\documentclass[preprint2]{aastex631}

\begin{document}

\title{JWST Near Infrared Spectroscopy of High Albedo Jupiter Trojans: A New Surface Type in the Trojan Belt} 

\correspondingauthor{Michael Brown}
\email{mbrown@caltech.edu}

\author[0000-0002-8255-0545]{Michael E. Brown}
\affiliation{Division of Geological and Planetary Sciences\\
California Institute of Technology\\
Pasadena, CA 91125, USA}
\author[0000-0001-9665-8429]{Ian Wong}
\affiliation{Space Telescope Science Institute, 3700 San Martin Drive, Baltimore, MD 21218, MD}

 \author[0000-0003-4778-6170]{Matthew Belyakov}
\affiliation{Division of Geological and Planetary Sciences\\
California Institute of Technology\\
Pasadena, CA 91125, USA}

\begin{abstract}
We present 0.8 to 5 $\mu$m JWST spectra of four $\sim$20 km diameter
Jupiter Trojans known to have albedos elevated above
the values
typical in the remaining Trojan population. 
The spectra of these four high albedo Jupiter Trojans are all similar,
with red slopes in the optical-IR transition region,
a break to lower slopes at 1.3 $\mu$m, and a broad
absorptions from 2.8 to 4 $\mu$m.  
The 0.8 to 2.5 $\mu$m spectra of these objects match
the spectra of neither the well-known
``red'' and ``less-red'' Jupiter Trojans nor of
any known asteroid taxonomic class.
The reflecticity of these objects does not rise
redward of 4 $\mu$m, 
a property
that is seen in the previous JWST observations of
Jupiter Trojans only in
Polymele. Indeed,
the high albedo Jupiter Trojan spectra
are a good match to that of Polymele, and Polymele is
both the smallest Jupiter Trojan in the previous 
JWST sample
and has the highest albedo of the objects 
in that sample.  We conclude that Polymele
and the other high albedo Jupiter Trojans 
represent a third class of Jupiter Trojans not
represented in the more heavily-studied larger 
objects and are perhaps the products of recent 
disruptions. The Lucy flyby of Polymele in September 2027
will give a direct view of one of this 
new class of Jupiter Trojans.
\end{abstract}

\keywords{}
\section{Introduction}
The Jupiter Trojan asteroids are a key population for understanding the dynamical
evolution of the early solar system. In the Nice family of dynamical instability 
models, the Jupiter Trojans are sourced from the same population as 
the Kuiper belt of the outer solar system and thus should share similar
initial compositions \citep{2005Natur.435..462M, 2010CRPhy..11..651M}. The surface compositions of the Kuiper belt objects (KBOs)
and the Jupiter Trojans, however, appear different. { KBOs
the size of Jupiter Trojans} have a wide range
of optical colors and have infrared spectra which show evidence of
water ice, CO$_2$, CO, and organics \citep{2012AJ....143..146B, 2011Icar..214..297B, 2023PSJ.....4..130B, 2024NatAs.tmp..103D}. Jupiter Trojans, in contrast, occupy
a narrower range in color \citep{2011AJ....141...25E, 2014AJ....148..112W}, and infrared spectra show only a shallow 3$\mu$m
absorption, a small signature of organics, and, in one known case, a small CO$_2$
absorption \citep{2016AJ....152..159B, 2024PSJ.....5...87W}. While these differences are significant, they could all be simple
signatures of the differences in surface insolation at the locations of
the Jupiter Trojans and that of the Kuiper belt. The interior compositions
of these objects remain unknown.

One solution to examining interior compositions is to look for collisional
fragments. In the Kuiper belt, fragments that are part of the Haumea collisional
family have surfaces of nearly pure water ice \citep{2007Natur.446..294B}, while in the main asteroid belt,
even the youngest
known collisional family in the main belt, the Karin family, at only $\sim$6 Myr old, has family members 
whose surfaces are indistinguishable from the background population
\citep{2009Icar..199...86H}.
At the Jupiter Trojans, the best-studied collisional family -- that of 
(3548) Eurybates -- has fragments that are slightly less-red than the rest of
the Jupiter Trojan population. Eurybates itself is the only
known member of the Jupiter Trojans to show an absorption feature due to CO$_2$ \citep{2024PSJ.....5...87W}, but the
reason for the presence of CO$_2$ is unclear, and otherwise family members
have no additional identifiable spectral
features \citep{2010Icar..209..586D}. The Eurybates family, however, appears ancient \citep{2022AJ....164..167M}, so
$\sim$1-4 Gyr of space weathering could have erased any spectral signatures of
the interiors \citep{1987JGR....9214933T, 2006ApJ...644..646B}.

Spectroscopy of the survivors of more recent collisions could still
retain interior spectral signatures. While current collisional models 
suggest that objects in the $\sim$1 km size range should be collisonally
active with collisional time scales of under 100 Myr \citep{2007A&A...475..375D}, a smaller number
of larger objects should also have undergone catastrophic
collision over the past 100 Myr. Such objects might reveal themselves
as having unusual high albedos compared to the rest of the populations.

\citet{2022AJ....164...23S} used a combination of thermal emission observations of
Jupiter Trojans from the Atacama Large Millimeter Array (ALMA) and 
the Wide-field Infrared Survey Explorer (WISE)  to robustly 
identify a small set of $\sim$ 20km Jupiter Trojans with albedos 
significantly elevated over the { $\sim0.05$ average value of
the remainder of the population} and suggested that these
could be collisional fragments of recent impacts with still-exposed 
subsurface materials. Here, we obtain 0.8-5.0 $\mu$m JWST spectra of
four members of this higher-albedo, { never previously-observed} population and use these spectra
to search for signatures of freshly exposed surface materials on these
objects.
\section{Observations}
Observations of four higher-albedo Jupiter Trojans were obtained with 
{ the Near Infrared Spectrograph (NIRSPEC)}
on JWST \citep{2022A&A...661A..80J} using the low-resolution PRISM mode and the integral field unit.
This observing mode provides a spectrum from 0.8-5.0 $\mu$m with a resolution
ranging from 30 to 400.
For each target, the { JWST-provided}
ephemeris position was sufficiently accurate that
we blindly acquired the targets into the $3\times 3$ arcsecond aperture
and performed a two-position dither, using a single integration in the NRSIRS2RAPID
and other observing conditions are given in Table 1.

\begin{deluxetable*}{cccccccccccc}
\tablecaption{JWST observational parameters}
\tablehead{\multicolumn{2}{c}{object} & \colhead{date} & \colhead {time} & \colhead{exp} & \colhead {r} & \colhead{$\Delta$} & \colhead{phase} & \colhead{diam} & \colhead{albedo} & \colhead{$\eta$} & \colhead{T$_{\rm max}$}}

\startdata
11488&	1998RM11&	2023-10-24&	19:21-19:29&  233 &5.01& 4.32& 9.0&	22.7& 0.09& 	0.77& 191\\
13331&	1998SU52&	2024-03-29&	18:50-18:57&  233 &4.70& 4.33& 11.9&	17.7& 0.13&	0.50& 218\\
42168&	2001CT13&	2024-04-21&	07:12-07:21&  350 &5.47& 5.40& 27.7&	18.1& 0.13&	0.78 & 181\\
18263&	Anchialos&	2024-05-10&	16:07-16:15&	  321& 5.07& 5.03& 27.4&	20.1& 0.10&	0.69 & 195\\
\enddata
\tablecomments{Table parameters include exposure time, in seconds, $r$, the heliocentric
distance, $\Delta$, the distance to JWST, $r$ (both in AU), diameter (km), the phase (Sun-asteroid-earth angle), in degrees, and albedo, derived from Simpson et al. (2022), and the beaming parameter, $\eta$, 
and maximum temperature, T$_{\rm max}$ in degrees K, obtained in this work.}

\end{deluxetable*}

After confirming the correct acquisition of the target and the successful
execution of the observations, we run the most recent JWST pipeline and calibration files\footnote{as of 12 July 2024} (version 1.14.0; jwst\_01252.pmap)
starting with the Step~1 ``\_uncal.fits" files. We perform a custom $1/f$ noise removal
on the derived ``\_rate.fits" data, as described in \citep{2023PSJ.....4..130B}, before running the
Step 2 pipeline. We extract spectra from the Step 2 ``\_s3d.fits" image cube products
using a custom  
point spread function (PSF)-fitting scheme. 
The PSF of the data is wavelength dependent
both because of standard diffraction effects and because of the
undersampling of the PSF by the detector pixels coupled
with the curvature of the spectrum across the detector.
This latter effect causes pixel-level PSF variations at
a characteristic wavelength scale of hundreds of pixels in the PRISM mode.
To mitigate this and other wavelength-dependent PSF changes,
we first perform a wavelength dependent sky subtraction by subtraction the median of the spatial pixels far removed from the target.
We next determine the PSF independently 
at each wavelength by taking a two-dimensional 
median of the data in a region
$\pm$50 pixels in the spectral dimension and normalizing the
result to an area of unity. The irradiance
is then found by fitting the data at each
wavelength to the empirically found PSF by using a weighted 
median -- where the weight of each pixel in the median is the
amplitude of the PSF at that positions -- of the ratio between the two-dimensional data and 
the PSF. This weighted-median scheme provides a robust 
method to account for the large number of bad pixels
on the NIRSPEC detector at the expense of the $\sim$25\% higher
noise associated with the use of a median rather than a mean. 
The two dither positions are averaged to obtain the
final one-dimensional spectra. Observational circumstances and 
properties of the target are shown in Table 1.

The spectra of all targets are shown in Figure 1. In addition, the spectrum of 
the star SNAP-2, an approximate solar analog \citep{Bohlin_2014}, obtained through Program 1128 and reduced
identically, is also shown. The majority
of the light from the Trojan targets is reflected sunlight, as seen 
by the similarity between the target spectra and those of the solar analog.
Beyond approximately 4.5 $\mu$m, thermal emission from the Jupiter Trojans 
begins to become apparent. 
\begin{figure}
\begin{center}
\hspace*{-2cm}\includegraphics[scale=.4]{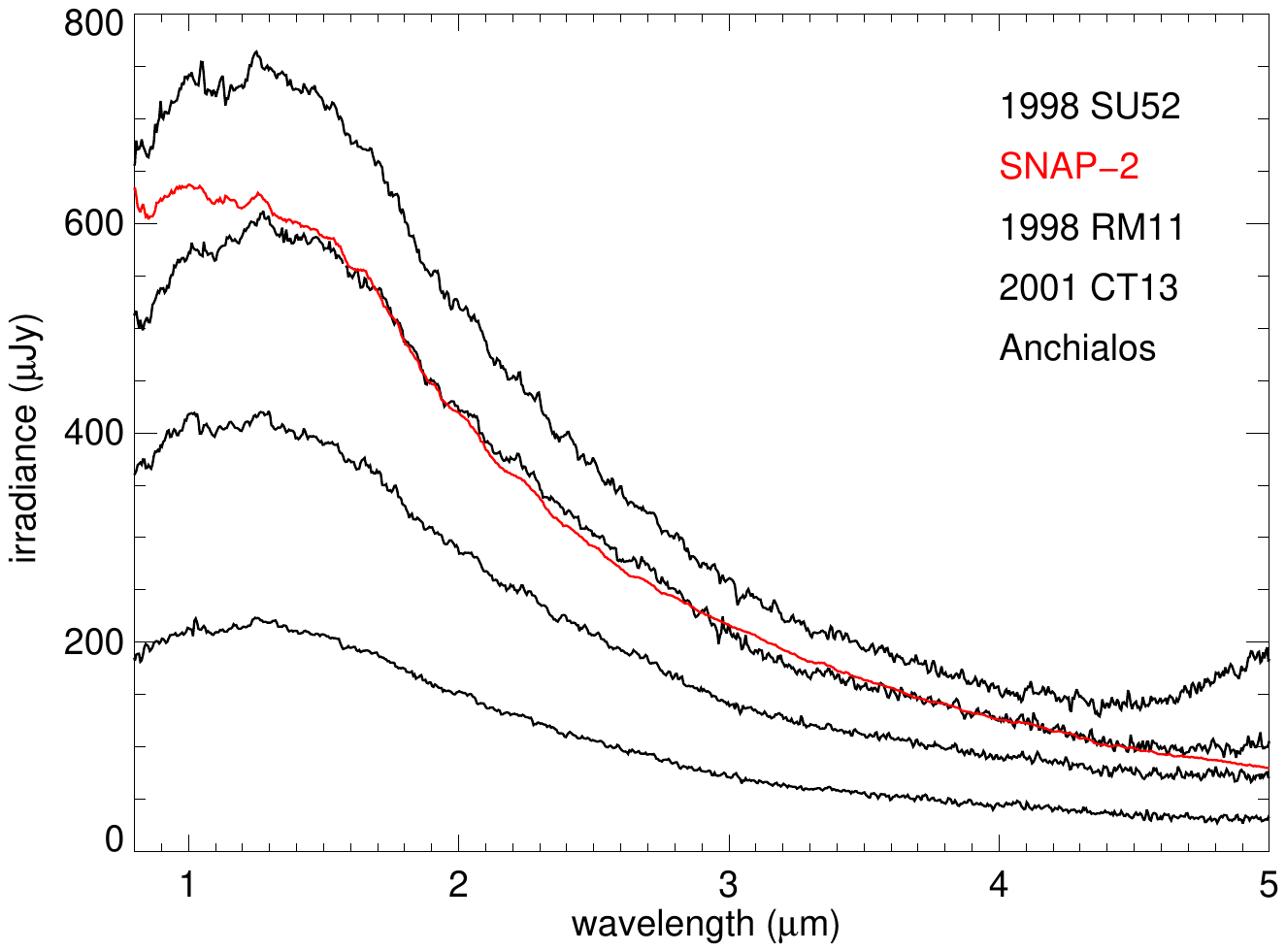}
    \caption{Spectra of four high albedo Jupiter Trojans (in black) along with the spectrum of SNAP-2,
    a solar analog star (in red). { The object name associated with
    each spectrum is shown in the figure in order from brightest to faintest
    at 1 $\mu$m.} The Trojan spectra are dominated by reflected sunlight except for
    the regions beyond $\sim$4.5 $\mu$m. { The relative signal-to-noise across
    the  spectra is determined by the total irradiance and the NIRSpec/PRISM
    throughput, both of which decrease to longer wavelength. The 
    total uncertainty can be estimated from the scatter in these largely featureless spectra.}}
    \end{center}
\end{figure}

\section{Results}
To examine the reflectance spectra of the targets, we must first
remove the contribution from thermal emission. We follow the procedure
outlined by \citet{2024PSJ.....5...87W}. In short, we use the ALMA-derived diameter
and geometric albedo of the targets along with the known geocentric and heliocentric distances
to create a model of emission based on the Standard Thermal Model \citep{1998Icar..131..291H},
with the beaming parameter, $\eta$, as a free parameter. 
We model a range
of beaming parameters, subtract the derived thermal emission
from the observed spectrum, and divide the resultant
spectrum by that of the solar analog star to obtain the
relative reflectance.
We then modify the beaming parameter until a satisfactory
fit is obtained. No rigorous prescription of what makes
a fit the most ``satisfactory'' is possible, as the true
reflectance spectrum is not known {\it a priori}. Based on
experimentation with the data, we find that assuming that the
reflectance continuum extends from the peak near 2.7 $\mu$m 
to a line through the spectrum between 3.8 and 5.0 $\mu$m 
and fitting the beaming parameter to minimize positive deviations
in this longer wavelength region gives consistent results.
Examples of this procedure are
shown in Fig 2. This procedure could artificially mask 
broad spectral features beyond about 4.2 $\mu$m.
We keep this caveat in mind when
examining the spectra of the objects.

\begin{figure}
\hspace*{-2cm}\includegraphics[scale=.5]{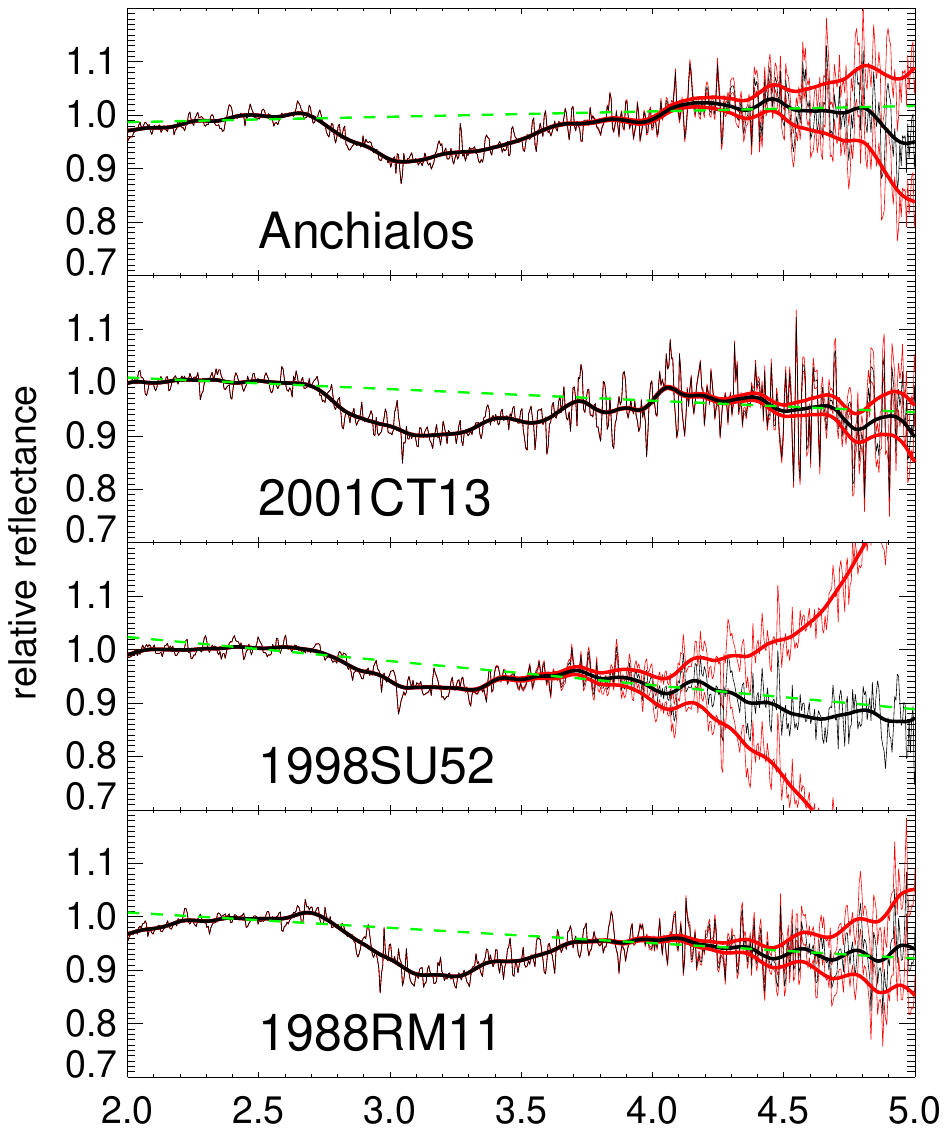}
   \caption{Relative reflectivity, scaled to unity between 2.5 and 2.7$\mu$m of the
   four high albedo asteroids with different choices for the beaming parameter. In each case
   we show our adopted beaming parameter as a black line and the derived reflectance for a 
   beaming parameter 0.1 higher (red curve below adopted reflectance) and one 0.05 lower
   (red curve above adopted reflectance). The green dashed line shows the extension from
   2.7$\mu$m to the region beyond 3.7$\mu$m, which we use as our estimate of
   the longer wavelength continuum.} 
\end{figure}

We calculate the absolute reflectivity of these objects by assuming that the
linear part of the spectrum from 0.8 to 1.4 $\mu$m extends linearly to 
0.55 $\mu$m at the center of the $V$ band where the albedo is defined. 
Such a linear slope in this region is characteristic of known Jupiter Trojans 
and many other small bodies with low albedos \citep{2011AJ....141...25E}.
Absolute reflectivities of the four high albedo Jupiter Trojan targets are shown in 
Figure 3. Other than slight changes in the absolute level of the reflectivites, the
spectra of all 4 targets are similar, with a spectral break at about 1.3 $\mu$m, 
similar to that on other Jupiter Trojans, and a broad absorption beyond 2.7 $\mu$m,
similar to that seen on previous JWST of lower albedo Jupiter Trojans \citep{2024PSJ.....5...87W}.

\begin{figure}
\hspace*{-2.5cm}\includegraphics[scale=.5]{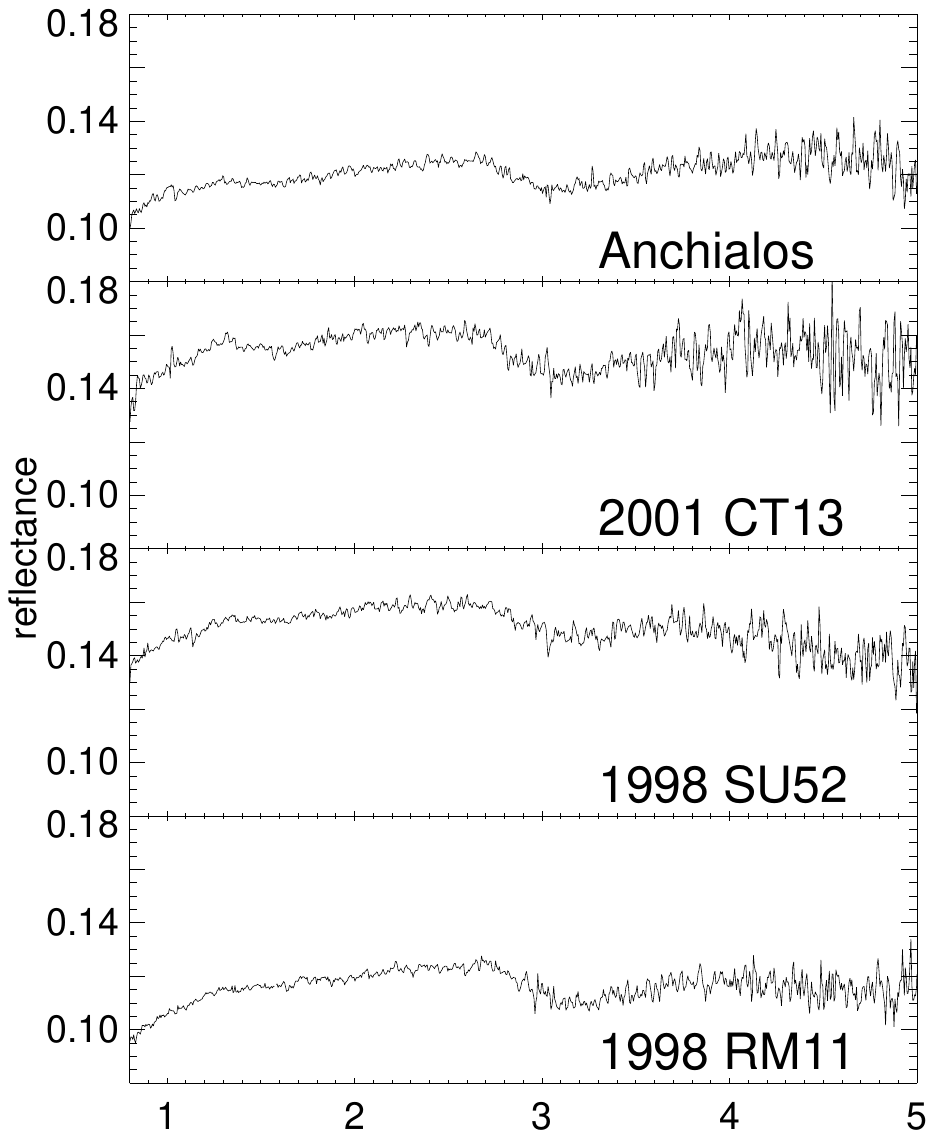}
\caption{Absolute reflectance of four high albedo Jupiter Trojans. The absolute reflectance
is determined by assuming that the linear spectrum from 0.8 to 1.3 $\mu$m extends downward
to 0.55 $\mu$m, where the visible albedo is defined.}
\end{figure}
\section{Discussion}
The spectra of the high albedo Jupiter Trojans do not contain
any immediately identifiable absorption features beyond the
3 $\mu$m feature noted by \citet{2016AJ....152..159B} and \citet{2024PSJ.....5...87W}.
In particular, no hint of water ice, which might be expected to
be exposed in a recent impact and which would lead to a
heightened albedo, is present. To demonstrate the lack
of water ice, in Figure 4, we show a simple
model consisting of a linear combination of two spectra.
{ The first component of our model is a spectrum of the large
Jupiter Trojan Eurybates 
\citep{2024PSJ.....5...87W} for which we extend the 
spectrum linearly below the observed wavelength. This
component has an overall albedo of 0.044 \citep{2023PSJ.....4...18M}. 
The second model
component consists of 10 $\mu$m grains 
of water ice, modeled using optical constants from \citet{2012Icar..218..831C} and using the technique of
\citet{1994Icar..108..243R}, and having an optical albedo
of 0.66. Typical water ice grain sizes for dark 
bodies at this heliocentric distance are
even larger \citep[i.e.][]{2020Icar..33713440S}, causing deeper absorption features, so our choice of small
grains is conservative. To obtain a surface albedo of
0.12, as is typical for these high albedo Trojans, we
require a mixture of 85\% Eurybates material and 15\%
of the water ice model.} 
We compare this model spectrum to an average of the
four high albedo Trojan spectra, where each spectrum has been scaled
to a value of unity at the median of the values between 2.5 and 2.7 $\mu$m, in Figure 4.
Such an ice exposure would be 
readily detected in these spectra. Very little water ice can be exposed on the surfaces
of these objects, and exposed water ice cannot account for the heightened albedos.
\begin{figure}
    \hspace*{-1.8cm}\includegraphics[scale=.4]{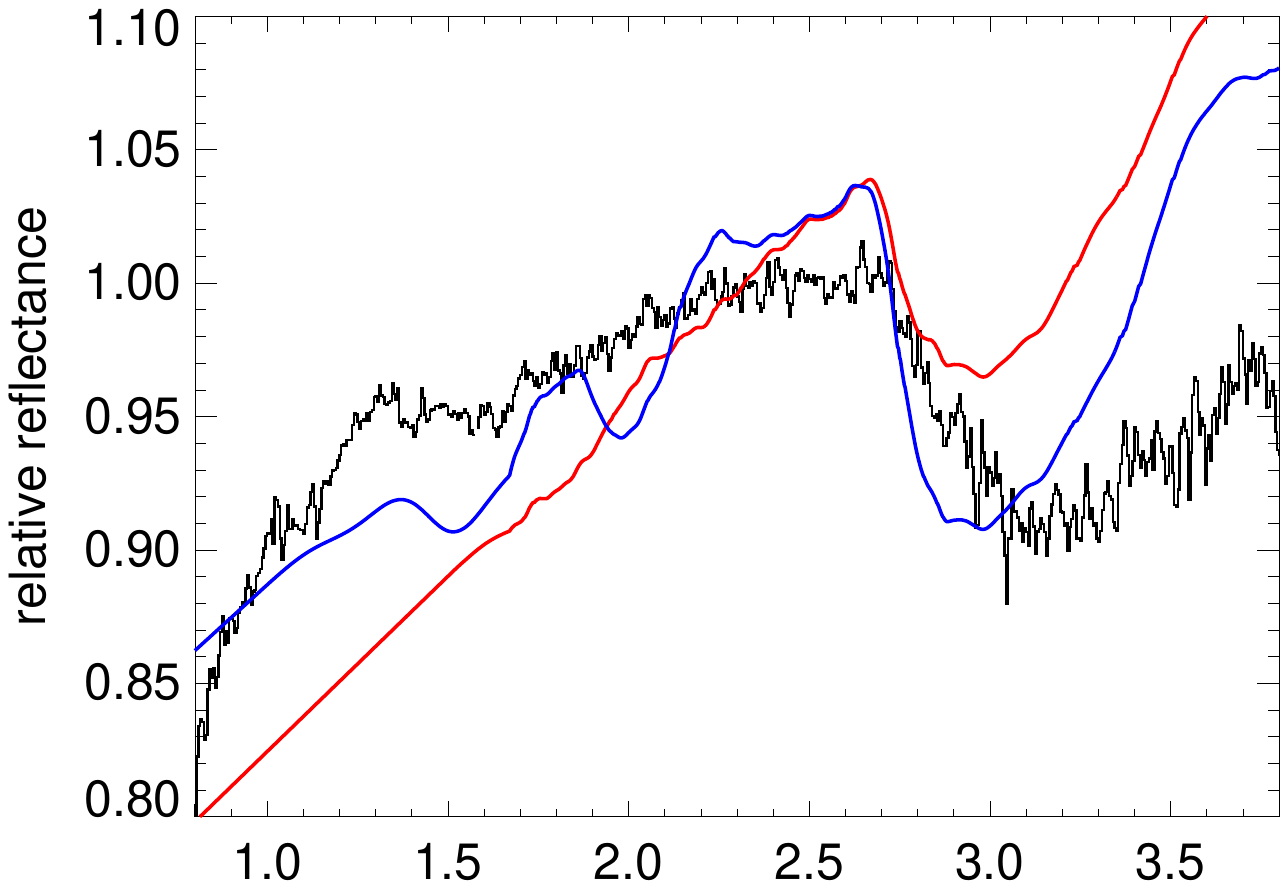}
    \caption{The average of the four high albedo Jupiter Trojans, scaled to unity at the median of the value between 2.5 and 2.7 $\mu$m (black), compared to a smoothed spectrum of Eurybates (red) and to a model consisting of 85\% coverage of Eurybates-like material and 15\% water ice (blue), giving an albedo similar to the high albedo Jupiter Trojans.}
\end{figure}

With no clear spectral signature of newly exposed materials,
we turn to comparison with the known spectral properties
of other Jupiter Trojans. In Fig. 5 we compare the average
high albedo Jupiter Trojan to the average
0.8 to 2.5 $\mu$m spectra of the ``red'' and ``less red''
classes of Jupiter Trojans from \citet{2011AJ....141...25E}.
The high albedo Trojans deviate significantly from either
class, with a spectral slope from 0.8 to 1.3 $\mu$m similar to 
that of the red class of Jupiter Trojans, but then a break to a 
much flatter spectrum than either the red or less-red Trojans
out to 2.5 $\mu$m. A possible broad absorption from 1.3 to
1.7 $\mu$m, unseen in previous Trojan spectra, is also
apparent. These feature can be seen, with varying strengths,
in all four of the individual spectra. This feature appears among
none of the known asteroid taxonomic classes \citep{2009Icar..202..160D}.

The spectral behaviour of the high albedo Jupiter Trojans beyond
2.5 $\mu$m is also unusual.
\citet{2016AJ....152..159B} observed 16 Jupiter Trojans from
2.3 to 3.8 $\mu$m using the Keck Observatory, while \citet{2024PSJ.....5...87W}  
observed 5 Trojans from 1.8 to 5.2 $\mu$m with JWST. The 
3~$\mu$m absorption in our high albedo Jupiter Trojan
targets has a similar shape and wavelength of those of the other Jupiter
Trojans, albeit the absorptions observed here are deeper and extend 
to longer wavelength than most of those previously
seen (Figure 5). These high albedo Trojans lack the 3.4 $\mu$m organic feature detected most 
strongly on (21900) Orus.
In addition they have no clear signature of the 4.26 $\mu$m
CO$_2$ absorption feature observed exclusively on Eurybates, to date \citep{2024PSJ.....5...87W}.

The largest difference between the high albedo Trojans and the majority of other previously observed
Trojans is that in the high albedo Trojans the continuum beyond 2.7 $\mu$m
is sloped
distinctly downward. Among the \citet{2024PSJ.....5...87W} JWST sample, only one object -- Polymele -- is similarly blue-sloped at the
longest wavelengths. Indeed, a comparison of Polymele with the average of the high albedo Trojans reveals
markedly similar spectral features throughout the range, with the 
high albedo Jupiter Trojans having an only slightly deeper 3 $\mu$m 
absorption than Polymele.
\begin{figure}
\hspace*{-1.5 cm}\includegraphics[scale=.4]{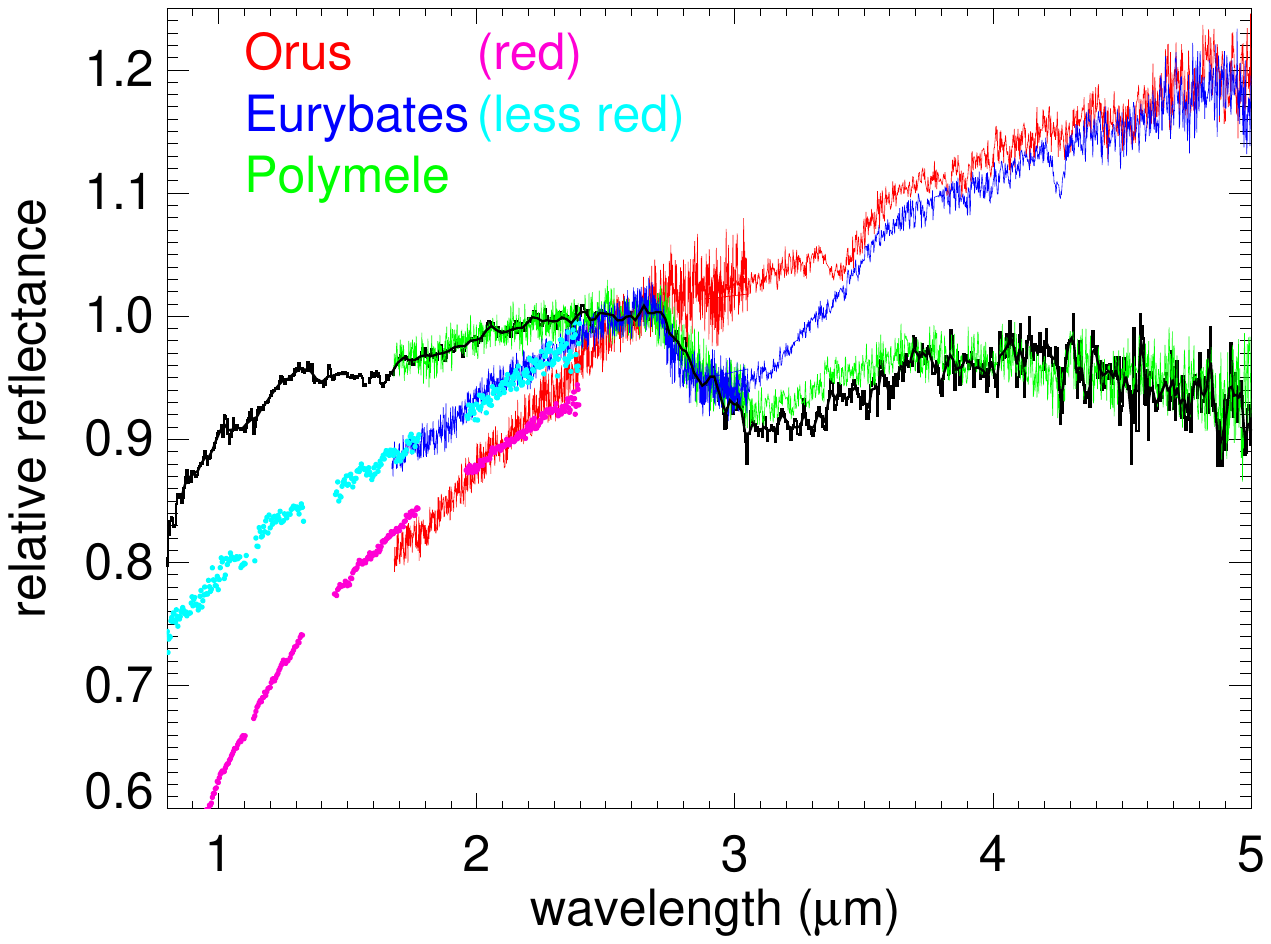}
\caption{Comparison of the average spectrum of the four high albedo 
Jupiter Trojans (black) to previous Trojan observations. 
The magenta points show the average of the red Trojan population 
from Emery et a. (2011) while the cyan points show the
average of the less-red population. The JWST spectrum
of Orus (red) -- a red
Jupiter Trojan -- shows a weak 3 $\mu$m absorption but a strong 3.4 $\mu$m
signature of organics, while Eurybates (blue) -- a less-red Trojan --
shows a strong
3 $\mu$m absorption and a distinct 4.26 $\mu$m signature of CO$_2$ (Wong et al. 2024). 
The high albedo Jupiter Trojans do not fit into either of
these categories.
Polymele is the highest albedo object in the Wong et al. (2024) sample
and has a spectrum remarkably similar to the other high albedo Jupiter Trojans.}
\end{figure}

Interestingly, Polymele is the highest albedo Jupiter Trojan in the 
\citet{2024PSJ.....5...87W}
JWST sample and the highest
albedo Trojan that will be visited by the Lucy Spacecraft. \citet{Buie_2018} give an albdo of 0.074 (compared to an average of $\sim$0.04 for the other Lucy targets), but continued occultation results have pushed the surface area even
smaller and thus the albedo even higher \citep{2023DPS....5510705L}.
Polymele is also the
smallest of the Wong et al. (2024) sample; at  $\sim$15 km 
it is approximately the same size as these high albedo
Jupiter Trojans. 

In addition to its unusually high albedo, Polymele has an optical color \citep{2020Icar..33813463S}
that places it ambiguously between the red and less-red
Jupiter Trojan colors \citep{2014AJ....148..112W} though
it is generally assigned to the class of less-red objects.
While the color measurements of the high albedo Jupiter Trojans
are not high quality, they are consistent with also being between
the two main Jupiter Trojan colors \citep{2021PSJ.....2...40S}.

The unusual spectral characteristics of these small high albedo Jupiter Trojans
and their similarities to Polymele suggests that these are
all members of a previously unrecognized class of Jupiter Trojans. 
This class appears to contain no objects larger than about $\sim$40 km,
while most are 20 km and smaller. These characteristics 
remain consistent
with the possibility that these are the rare larger objects that have
had recent large-scale collisions \citep{2022AJ....164...23S}, but no obvious spectral
signatures of such collisions are apparent. \citet{2015AJ....150..174W} have
argued that the collisionally active smallest Jupiter Trojans 
are consistent in color with the less-red population, inconsistent with
the idea that these high albedo objects with mid-red colors are 
products of collisions. While the true nature of these objects is unclear,
the Lucy flyby of Polymele in September 2027 and the comparison with 
the later flyby targets will give a rare opportunity
to help to answer some of the mysteries of these high albedo Jupiter Trojans.

\begin{acknowledgments}
We thank Jessica Sunshine and Rick Binzel for sharing
interesting insights into these observations. This work is based on observations made with the NASA/ESA/CSA James Webb Space Telescope. The data were obtained from the Mikulski Archive for Space Telescopes (MAST) at the Space Telescope Science Institute, which is operated by the Association of Universities for Research in Astronomy, Inc., under NASA contract NAS 5-03127 for JWST. These observations are associated with program \#2869. Support for program \#2869 was provided by NASA through a grant from the Space Telescope Science Institute, which is operated by the Association of Universities for Research in Astronomy, Inc., under NASA contract NAS 5-03127. The specific observations analyzed can be accessed via \dataset[DOI: 10.17909/j5w3-0d36]{https://doi.org/10.17909/10.17909/j5w3-0d36}.
\end{acknowledgments}
\facilities{JWST/NIRSpec.}

\eject

\end{document}